\documentclass{pazha2}
\usepackage{epsf}
\usepackage{graphicx}
\usepackage{color}
\usepackage{multirow}
\definecolor{darkblue}{rgb}{0,0,0.9}
\def\*{$^{*}$}
\def\a{$^{\mbox{\small a}}$}
\def\bb{$^{\mbox{\small b}}$}
\def\cc{$^{\mbox{\small c}}$}
\def\dd{$^{\mbox{\small d}}$}
\def\e{$^{\mbox{\small e}}$}
\def\f{$^{\mbox{\small f}}$}

\begin{document}
\journalinfo{2018}{44}{6}{378}{413}{425}[389]
\sloppypar
%\begin{flushleft}
%{\it to be published in Astronomy Letters, 2018, v. 44, n. 6,
%  pp. 378--389}\\
%{\it (in Russian: Pis'ma v Astronomicheskii Zhurnal, 2018,
%     v. 43, No. 10, pp. 413-425)}\\ [30mm]
%\end{flushleft}

\title{Low-Frequency Quasi-Periodic Oscillations in the
  X-ray Nova MAXI\,J1535-571 at the Initial Stage of Its 2017 Outburst}

\author{I. A.~Mereminskiy\email{i.a.mereminskiy@gmail.com;~grebenev@iki.rssi.ru}\addres{0},
  S. A.~Grebenev{\textsuperscript{*}}\addres{0}, A. V.~Prosvetov\addres{0}, and
  A. N.~Semena\addres{0}    
 \addrestext{0}{Space Research Institute, Russian Academy of Sciences,
   Profsoyuznaya ul. 84/32, Moscow, 117997 Russia}
}
\shortauthor{MEREMINSKIY et al.}
\shorttitle{LOW-FREQUENCY QUASI-PERIODIC OSCILLATIONS IN MAXI\,J1535-571}

%\received{\today}
\submitted{December 1, 2017}
%\received{December 1, 2017}
%\revised{~}

\begin{abstract}
  \noindent
We report the discovery of low-frequency quasi-periodic
oscillations (QPOs) in the power spectrum of the X-ray nova
MAXI\,J1535-571 at the initial stage of its outburst in
September 2017. Based on data from the SWIFT and INTEGRAL
instruments, we have traced the evolution of the QPO parameters
(primarily their frequency) with time and their correlation with
changes in the X-ray spectrum of the source (changes in the
emission flux and hardness). We place constraints on the
theoretical QPO generation models.\\

\noindent
{\bf DOI:} 10.1134/S106377371806004X\\

\end{abstract}
%\vfill
%\noindent\rule{8cm}{1pt}\\
%{$^*$ e-mail $<$i.a.mereminskiy@gmail.com$>$}
%\clearpage

%*************************************************************
\section*{INTRODUCTION}
\noindent
The X-ray transient MAXI\,J1535-571 with a hard power-law
emission spectrum was discovered on September 2, 2017, in the
constellation Norma, $\sim$\,1\deg\ away from the Galactic
plane, simultaneously by the GSC/MAXI instrument (Negoro et
al. 2017) and the BAT/SWIFT monitor (Markwardt et al. 2017).
The promptly performed observations by the XRT and UVOT
telescopes of the SWIFT observatory allowed the source to be
localized with an accuracy up to 1\farcs5 (Kennea et al. 2017)
and then its optical counterpart to be found (Scaringi et
al. 2017). The source was detected in the radio (Russell et
al. 2017) and near-infrared (Dincer 2017) bands as well as at
millimeter wavelengths (Tetarenko et al. 2017).  An additional
confirmation that the optical object is actually the counterpart
of the X-ray one was the presence of the Br${\gamma}$ hydrogen
line (Britt et al. 2017) in its infrared spectrum, which is
associated with accretion processes (Bandyopadhyay et al. 1997).
One week after the discovery of the source, Nakahira et
al. (2017) and Kennea (2017) reported a softening of its X-ray
spectrum.\\ [-3.5mm]

Such a behavior is typical for X-ray novae, i.e., low-mass X-ray
binaries in which there is nonstationary accretion onto a black
hole (or a neutron star with a weak magnetic field). It is
common practice to describe the development of the outbursts of
such binaries in terms of the change of their ``states'' caused
by a change in the regime of accretion. During each state the
X-ray nova emission has its unique spectral and timing
properties (see, e.g., Sunyaev et al. 1988, 1991; Grebenev et
al. 1993, 1997; Tanaka and Shibazaki 1996; Remillard and
McClintock 2006; Belloni et al. 2010; Cherepashchuk 2013).
%----------------------------------------------------------------------------------------------
\begin{table*}[tp]
\vspace{5mm}
\centering
{{\bf Table 1.} Observations and parameters of the LF QPOs in \mbox{MAXI\,J1535-571}\protect\\}
\label{tab:qpo}
\vspace{4mm}
%\small
\begin{tabular}{c|c|c|c|c|c|c@{}} \hline\hline
Instrument\a & Observation & Beginning & End &
\multicolumn{2}{c|}{QPOs}&1H\bb\\ \cline{2-6}
& & \multicolumn{2}{c|}{} & & \\ [-3.3mm]
 &ID  &\multicolumn{2}{c|}{MJD-58000.00} & Frequency\cc & Width\dd&\\  
\hline
& &&& && \\ [-3.3mm]
XRT+BAT& 10264003        &4.28 &4.36&  $0.21\pm 0.03 $&$0.22\pm 0.07$&*\e\\
IBIS         & 18600002-06   &4.53 &4.75&  $0.23\pm 0.01 $&$0.14\pm0.06$\\
~BAT\f   & 39407002        &4.63 &4.93&  $0.24\pm 0.01 $&$0.14\pm0.06$\\
IBIS         & 18600007-11   &4.75 &4.96&  $0.24\pm 0.01 $&$0.14\pm0.04$\\
IBIS         & 18600012-16   &4.96 &5.18&  $0.28\pm 0.02 $&$0.16\pm0.08$\\
IBIS         & 18600017-21   &5.18 &5.39&  $0.37\pm 0.01 $&$0.11\pm0.04$\\
~BAT\f   & 87473001         &5.28 &5.45&  $0.37\pm 0.01 $&$0.11\pm0.04$\\
IBIS         & 18600022-26    &5.39 &5.63&  $0.45\pm 0.01 $&$0.15\pm0.04$\\
IBIS         & 18600027-31    &5.63 &5.88&  $0.53\pm 0.01 $&$0.11\pm0.03$\\
IBIS         & 18600032-37    &5.89 &6.14&  $0.68\pm 0.02 $&$0.19\pm0.05$\\
IBIS         & 18600038-42    &6.14 &6.36&  $0.89\pm 0.02 $&$0.16\pm0.04$\\
IBIS         & 18600043-47    &6.36 &6.57&  $1.03\pm 0.01 $&$0.13\pm0.03$\\
~BAT\f   & 30806051         &6.27 &6.95&  $1.03\pm 0.01 $&$0.13\pm0.03$\\ 
XRT+BAT& 10264004         &7.27 &7.29&  $1.87\pm 0.03 $&$0.21\pm 0.05$&*\\
XRT         & 10264005         &8.26 &8.27&  $2.15\pm 0.02 $&$0.10\pm 0.03$&*\\
XRT+BAT& 10264007         &9.01 &9.02&  $2.67\pm 0.03 $&$0.26\pm 0.09$&\\
JEM-X      & 18620019         &10.54&10.58& $1.98\pm 0.05 $&&\\
XRT+BAT& 10264006         &10.93&10.94&  $2.22\pm 0.03 $&$0.20\pm 0.05$&*\\
JEM-X      & 18620034         &11.19&11.23 & $2.11\pm 0.06 $&&\\
XRT+BAT& 10264008         &11.39&11.40&  $2.32\pm 0.03 $&$0.24\pm 0.08$&*\\
XRT+BAT& 10264009         &12.12&12.13&  $2.29\pm 0.02 $&$0.19\pm 0.06$&*\\
XRT         & 88245001         &13.18&13.20&  $2.59\pm 0.02 $&$0.24\pm 0.05$&*\\
XRT         & 88245001         &13.24&13.25&  $2.89\pm 0.06 $&$0.32\pm 0.18$&\\
JEM-X      & 18630034         &13.89&13.92& $2.64\pm 0.04 $&&\\ 
XRT+BAT& 10264010         &14.18&14.19&  $3.22\pm 0.04 $&$0.29\pm 0.07$&*\\
XRT         & 88245002         &16.97&16.98&  $-$&&\\
\hline
\multicolumn{7}{l}{}\\ [-2mm]
\multicolumn{7}{l}{\a\ The XRT and BAT telescopes of the SWIFT observatory and the JEM-X and }\\
\multicolumn{7}{l}{\ \ \  IBIS/ISGRI of the INTEGRAL observatory.}\\
\multicolumn{7}{l}{\bb\ The presence of the first harmonic of the QPO peak in the power spectrum.}\\
\multicolumn{7}{l}{\cc\ The frequency of the Lorentzian peak, Hz.}\\
\multicolumn{7}{l}{\dd\ The width of the Lorentzian peak, Hz; it depends on energy and, therefore, the }\\
\multicolumn{7}{l}{\ \ \ peak is narrower in the IBIS spectrum.}\\
\multicolumn{7}{l}{\e\  In this spectrum the first harmonic at
  $f_{\rm 1H}=0.42\pm 0.02$ Hz is noticeably more}\\
\multicolumn{7}{l}{\ \ \ significant than the fundamental one.}\\
\multicolumn{7}{l}{\f\ BAT was used only for the spectral measurements, the QPO frequency was }\\
\multicolumn{7}{l}{\ \ \ determined from the simultaneous IBIS observations.}\\
\end{tabular}
\vspace{-6mm}
\end{table*}
%----------------------------------------------------------------------------------------------
%----------------------------------------------------------------------------------------------
\begin{figure*}[t]
\centerline{\includegraphics[scale=0.76]{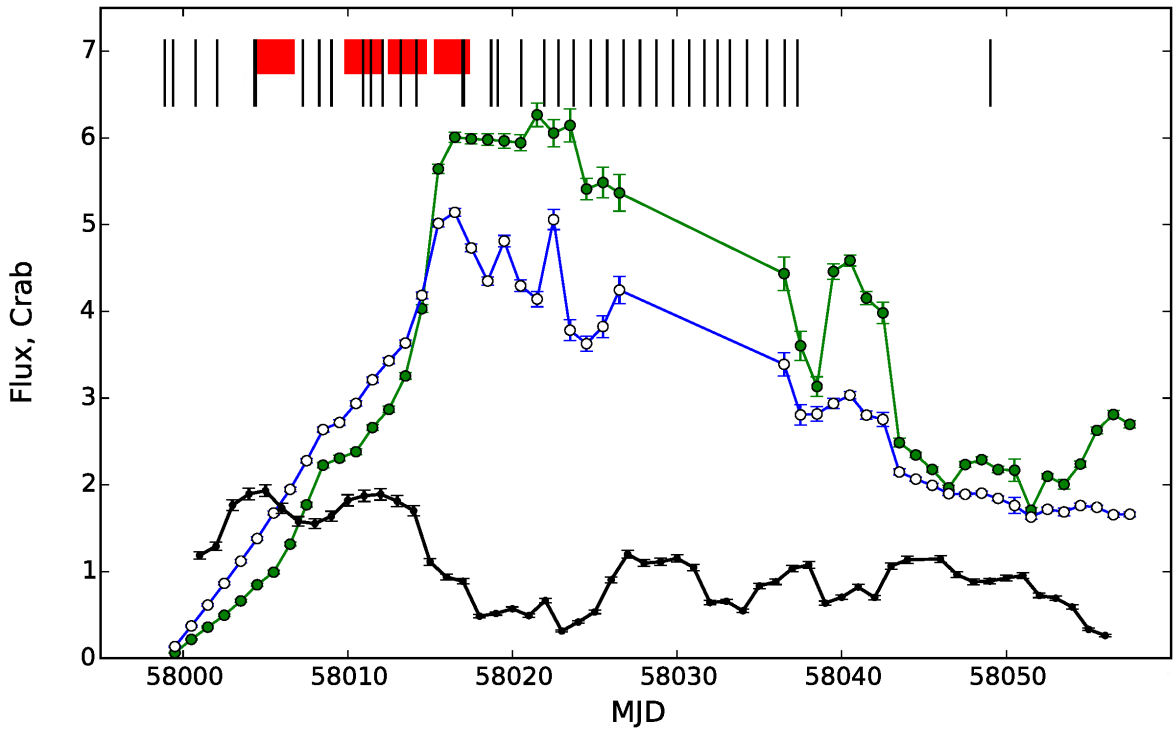}}
\caption{Light curves of the X-ray nova MAXI\,J1535-571. The
  black small circles indicate the 15--50 keV flux from the
  BAT/SWIFT data, the green and blue (filled and open) large
  circles indicate the 2--4 and 4--10 keV fluxes from the MAXI
  data. MJD\,58000.00 corresponds to September 4, 2017. The red
  broad lines and black narrow lines at the top specify the
  intervals of INTEGRAL and XRT/SWIFT observations,
  respectively.}
\label{fig:lc}
\end{figure*} 
%----------------------------------------------------------------------------------------------

All outbursts of X-ray novae begin with a low (hard) state in
which the central, hot, optically thin accretion disk region
puffed up by instabilities makes a major contribution to the
radiation. The observed power-law spectrum with an exponential
cutoff at high energies is formed in this region by
Comptonization of the radiation on high-temperature electrons.
As the luminosity (accretion rate) increases, the hard state
changes into an intermediate hard one, then an intermediate soft
one, and, finally, a high (soft) state.  The radiation from the
outer cold, optically thick accretion disk begins to dominate in
the source's spectrum. The transition between the states
suggests that the size of the inner high-temperature region
decreases with increasing accretion rate. At such an accretion
rate the outer disk becomes too dense and its evaporation begins
later, closer to the black hole, in the region of maximum energy
release.\\ [-2mm]

Measuring the radius of the inner region (or the blackbody disk
truncation radius) $r_{\rm tr}$ at different outburst stages is
critical for testing this picture.  However, the results of such
measurements are so far contradictory --- an analysis of the
blackbody component in the spectrum of X-ray novae shows that in
a number of systems the cold accretion disks with $kT_{\rm
  br}\sim0.1$--$0.5$ keV probably reach the innermost stable
circular orbit near the black hole (Miller et al. 2006a, 2006b;
Reis et al. 2011), while an analysis of the hard X-ray component
reflected from the cold disk and, in particular, the
relativistically broadened 6.4 keV neutral iron line leads to
large truncation radii in some cases (F\"{u}rst et al. 2015) and
to small ones in other cases (Miller et al. 2015).

The different states of X-ray novae differ not only by the
spectral shape but also by the pattern of their rapid flux
variability (Grebenev et al. 1993; Belloni 2010). In the low
(hard) state the X-ray power spectrum usually includes a broad
frequency limited (``red-white'') noise component on which one or
more narrow Lorentzian peaks with frequencies $f_{\rm qpo}$ from
$\sim0.1$ to several Hz indicative of low-frequency
quasi-periodic oscillations of the flux (LF QPOs)
are superimposed. Such LF QPOs were first
described by Ebisawa et al. (1989) and Grebenev et al. (1991).
The red-white noise can be successfully fitted
by a King function. It has been established that the break
frequency $f_{\rm br}$ above which the noise amplitude begins to
rapidly decrease is related by a linear relation to the
fundamental QPO frequency $f_{\rm qpo}$ (Wijnands and van der
Klis 1999; see also Prosvetov and Grebenev 2015), with this
relation also holding for both systems with black holes and
systems with neutron stars.

The origin of LF QPOs has not yet been established, despite the
numerous attempts to do this (see, e.g., Vikhlinin et al. 1994;
Stella and Vietri 1998; Titarchuk et al. 2007; Ingram et
al. 2009). It is important that in some QPO models, for example,
relativistic disk precession (Stella and Vietri 1998) or
diffusive propagation of perturbations in the disk (Titarchuk et
al. 2007), the frequency $f_{\rm qpo}$ directly depends on the
truncation radius of the accretion disk $r_{\rm tr}$, which
allows the change in this radius (the size of the
high-temperature central zone) during the outburst to be traced.

Such LF QPOs with a gradually changing frequency were detected
in the X-ray power spectrum of \mbox{MAXI\,J1535-571} on September 11,
2017, eight days after its discovery (Mereminskiy and Grebenev
2017).  This paper is devoted to their detailed analysis.

%*************************************************************
\section*{DATA ANALYSIS}
\noindent
After the discovery of MAXI\,J1535-571, its observations were
begun at many of the then operating Xray telescopes. In this
paper we use the observations of this source by the IBIS/ISGRI
(Lebrun et al. 2003; Ubertini et al. 2003) and \mbox{JEM-X}
(Lund et al. 2003) telescopes of the INTEGRAL observatory
(Winkler et al. 2003), the BAT (Barthelmy et al. 2005) and XRT
(Burrows et al. 2005) telescopes of the SWIFT observatory
(Gehrels et al. 2004), and the GCS monitor (Mihara et al. 2011)
of the MAXI observatory (Matsuoka et al. 2009).
%----------------------------------------------------------------------------------------------
\begin{figure}[ht]
\centerline{\includegraphics[width=1.05\linewidth]{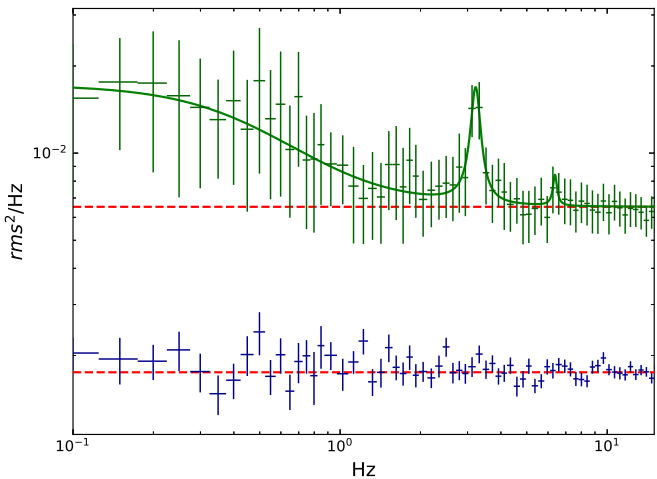}}
\caption{XRT/SWIFT power spectra for MAXI\,J1535-571 in the
  energy range 0.8--10 keV. The upper (green dots, the spectrum
  was multiplied by 4 for clarity) and lower (blue dots) spectra
  represent, respectively, the observations on MJD\,58014,
  before the transition to the soft state, and on MJD\,58017, after
  the transition. Low-frequency broadband noise and two QPO
  peaks corresponding to the fundamental frequency and the first
  harmonic are clearly seen in the power spectrum before the
  transition. There is virtually no noise in the power spectrum
  after the transition.}
\label{fig:powtrans}
\end{figure}
%----------------------------------------------------------------------------------------------

The list of observations of the source by the INTEGRAL and SWIFT
instruments used in the paper is given in Table\,1. Note that a
powerful solar flare occurred at the beginning of revolution
1861 of the INTEGRAL satellite (MJD\,58007.3-58009.4), as a
result of which the observations planned for this revolution
were interrupted.

%*************************************************************
\subsection*{Investigation of the Variability}	
\noindent
We were interested primarily in the rapid variability of the
source. The XRT/SWIFT data obtained in the fast timing mode were
passed through the standard pipeline \texttt{xrtpipeline} and
then barycentered.  Because of the source's brightness, the
telescope performed most of the observations in the regime of
``overloading''. In this case, the photon arrival frequency
exceeded the detector array sampling frequency.  If several
photons fell on the sampled detector area between readouts, then
they either were recorded as one higher-energy photon or were
ignored altogether, because their total energy exceeded the
specified threshold. When analyzing the temporal variability in
this regime, we, however, did not resort to a standard practice,
i.e., the exclusion of detector areas with an excessively large
count rate. Overloading affects primarily the measured energy
spectrum of the source and has no significant influence on the
QPO frequency, whose measurement was our most important task. We
made sure of this in several observations by excluding five most
illuminated detector array columns in each of them. For our
analysis we used the light curves with a resolution of 20 ms in
the energy range 0.8--10 keV.

The data from the JEM-X and IBIS/ISGRI telescopes of the
INTEGRAL observatory were processed with the standard data
analysis package \texttt{OSA-10} and barycentered. We used the
light curves with a resolution of 0.1 s in the energy ranges
3--20 and 20--200 keV, respectively. Several X-ray sources fell
within the fields of view of the telescopes at once.  Although we
used the detector pixel illumination fraction (PIF) by photons
from the source of interest (MAXI\,J1535-571) when constructing
its light curves, the photons from other sources also made a
certain contribution. Fortunately, MAXI\,J1535-571 was much
brighter than other objects and the distortions introduced by
them were insignificant.
%----------------------------------------------------------------------------------------------
\begin{figure*}[t]
\centerline{\includegraphics[width=0.84\linewidth]{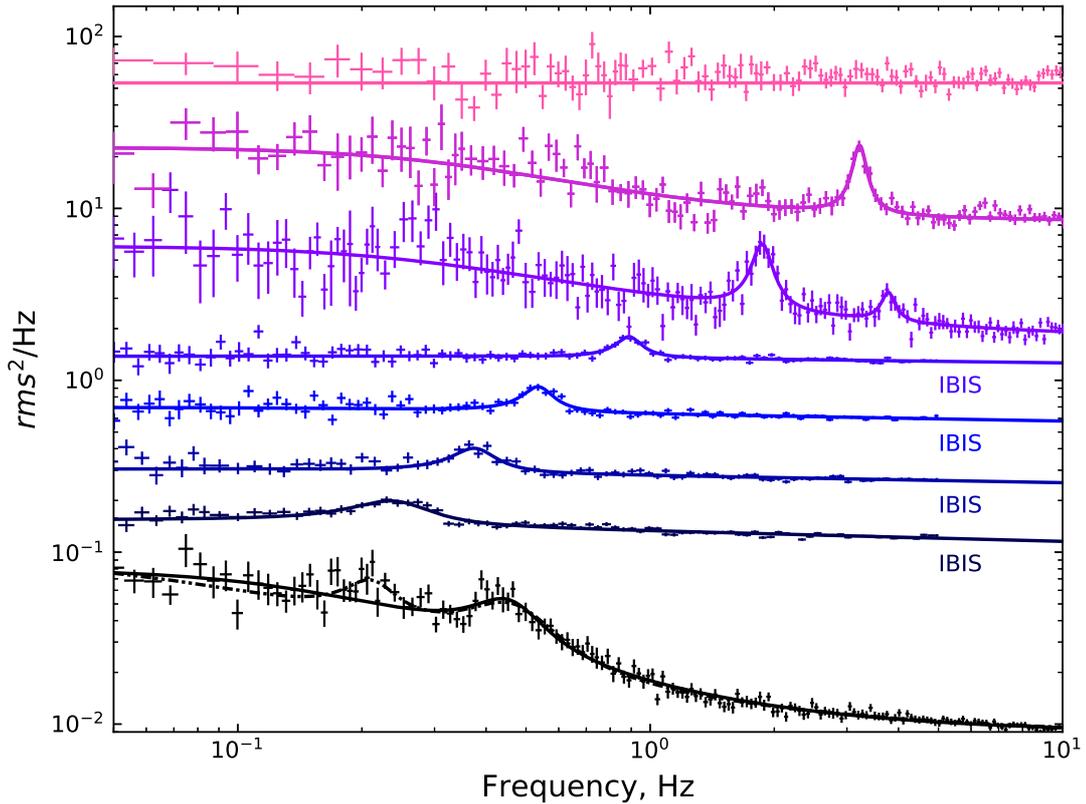}}
\caption{Evolution of the X-ray power spectrum and QPO peak
  (peaks) for MAXI\,J1535-571 from the XRT/SWIFT and
  IBIS/ISGRI/INTEGRAL data in the period MJD\,58004--58017. The
  spectra in observations 10264003, 1860002-06, 18600017-21,
  18600027-31, 18600038-42, 10264004, and 88245002, which are
  normalized to some factor for clarity, are shown from bottom
  to top. The solid lines indicate the best fits to the spectra
  by the adopted model (see the text). In the early spectrum the
  fundamental peak is indicated by the dashed line, because its
  significance is much lower than that of the first
  harmonic. The shift of the QPO peak toward higher frequencies
  from $f_{\rm QPO}\sim0.23$ Hz to $\sim3.22$ Hz and then its
  disappearance in observation on MJD\,58017, after the source's
  transition to the soft state, are clearly seen.}
\label{fig:qpodrift}
\end{figure*}
%----------------------------------------------------------------------------------------------

%*************************************************************
\subsection*{Investigation of the Radiation Spectra}
\noindent
The evolution of the source's radiation spectrum during its
outburst is also of interest. To obtain broadband (2--150 keV)
spectra, we used the quasi-simultaneous MAXI and BAT/SWIFT
observations.  We took the parts of the spectra in the energy
range 2--20 keV from the web page of the MAXI monitor
(\texttt{http://maxi.riken.jp/mxondem}) and the parts of the
spectra in the harder energy range 20--150 keV from the
BAT/SWIFT observations in the survey mode. We used the
histograms of detector events accumulated during the sky survey
with a typical exposure time of 0.5--1.5 ks. Since such data are
analyzed quite rarely, we described in detail the spectrum
extraction procedure in the Appendix.
%----------------------------------------------------------------------------------------------
\begin{figure*}
\centerline{\includegraphics[width=0.75\linewidth]{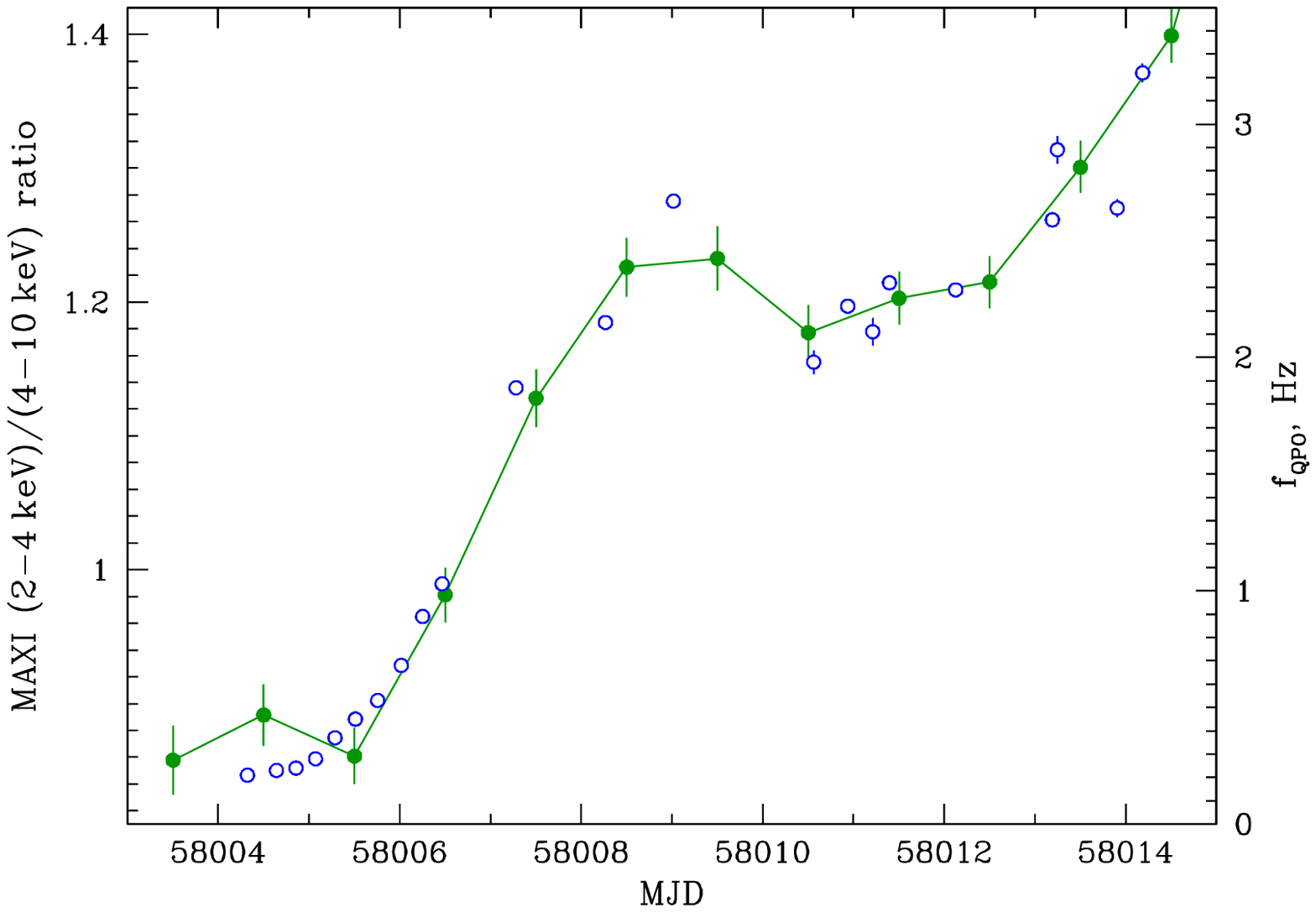}}
\caption{Evolution of the QPO frequency during the rising phase
  of the outburst from the data of Table\,1 (open/blue
  circles). The filled (green) circles connected by the solid
  line indicate the change in the source's X-ray softness (the
  ratio of the fluxes in the 2--4 and 4--20 keV subbands from
  the MAXI data).}
\label{fig:qpoevol}
\end{figure*}
%----------------------------------------------------------------------------------------------

%*************************************************************
\section*{RESULTS}
\subsection*{Light Curves and the Transition to the Soft State}	
\noindent
Figure~\ref{fig:lc} presents the source's light curves during
its outburst in the soft (2--4 and 4--10 keV, from the MAXI
data) and hard (15--50 keV, from the BAT/SWIFT data) X-ray
bands. The rising phase lasted for approximately 16 days (until
MJD\,58015) in the soft X-ray band and was much shorter (about 5
days) in the hard one. Near MJD\,58015 the hard X-ray flux
dropped sharply (on a time scale of 2 days) by a factor of
$\sim3$, while the soft one rose by a factor of $\sim2$, with
the 2--4 keV flux having exceeded the 4--10 keV flux for the
first time. At this moment the spectral state of the source
changed from hard to soft. It remained such at least for the
next one and a half months, despite the fact that, as follows
from the figure, the soft X-ray flux began to gradually decrease
after $\sim 10$ days of relative stability and reached the level
observed in the hard state by the end of the observations.

Simultaneously with the transition of the source to the soft
spectral state, the shape of its X-ray power spectrum also
changed. Figure\,\ref{fig:powtrans} presents the power spectra
obtained during the XRT/SWIFT observations of the source at
MJD\,58014 and 58017 (ObsID\,10264010 and 88245002,
respectively). The first spectrum has a shape with pronounced
powerful low-frequency noise (LFN) and two QPO peaks typical for
the hard state, while the second spectrum is virtually
indistinguishable from a flat ``white'' noise spectrum and does
not contain any features, which is typical for the soft state of
X-ray novae. In the energy range 0.8--10 keV the power of LFN
(in the frequency range 0.1--10 Hz) and the fundamental QPO peak
was $\simeq2$\% and $4$\% of the photon flux in the source's
hard state and in total less than $2.7$\% in its soft state
(90\% confidence interval). Note that the feature in the light
curves near MJD\,58008 resembling the previous, failed attempt
of the transition to the soft state was also accompanied by
pronounced changes in the source's power spectrum.

As has already been said, this paper is devoted to investigating
the quasi-periodic oscillations of the flux from the
source. Therefore, below we will restrict our analysis to the
initial outburst stage, i.e., before the source's transition to
the soft (high) state.

%*************************************************************
\subsection*{Quasi-Periodic Oscillations}
\noindent
We analyzed all the accessible XRT/SWIFT observations of the
source performed in the timing mode. If an observation included
several exposures separated by long interruptions, then the time
intervals corresponding to them were considered separately. The
power spectrum of each observation was fitted by a combination
of the King function describing the broadband noise, one or two
Lorentzians describing the QPOs, and the constant responsible
for the Poisson noise power. Examples of our power spectra and
their best fits by the adopted model at different outburst
stages are shown in Fig.~\ref{fig:qpodrift}. The measured
frequencies and widths of the fundamental QPO peak are given in
Table\,1. The asterisks mark the observations in which the first
harmonic of the QPO peak was recorded. Curiously, during the
first XRT/SWIFT observation of the source (MJD\,58004.3) the
first harmonic in the power spectrum was brighter and
appreciably more significant than the fundamental QPO peak.
%----------------------------------------------------------------------------------------------
\begin{figure*}[th]
\centerline{\includegraphics[width=0.8\linewidth]{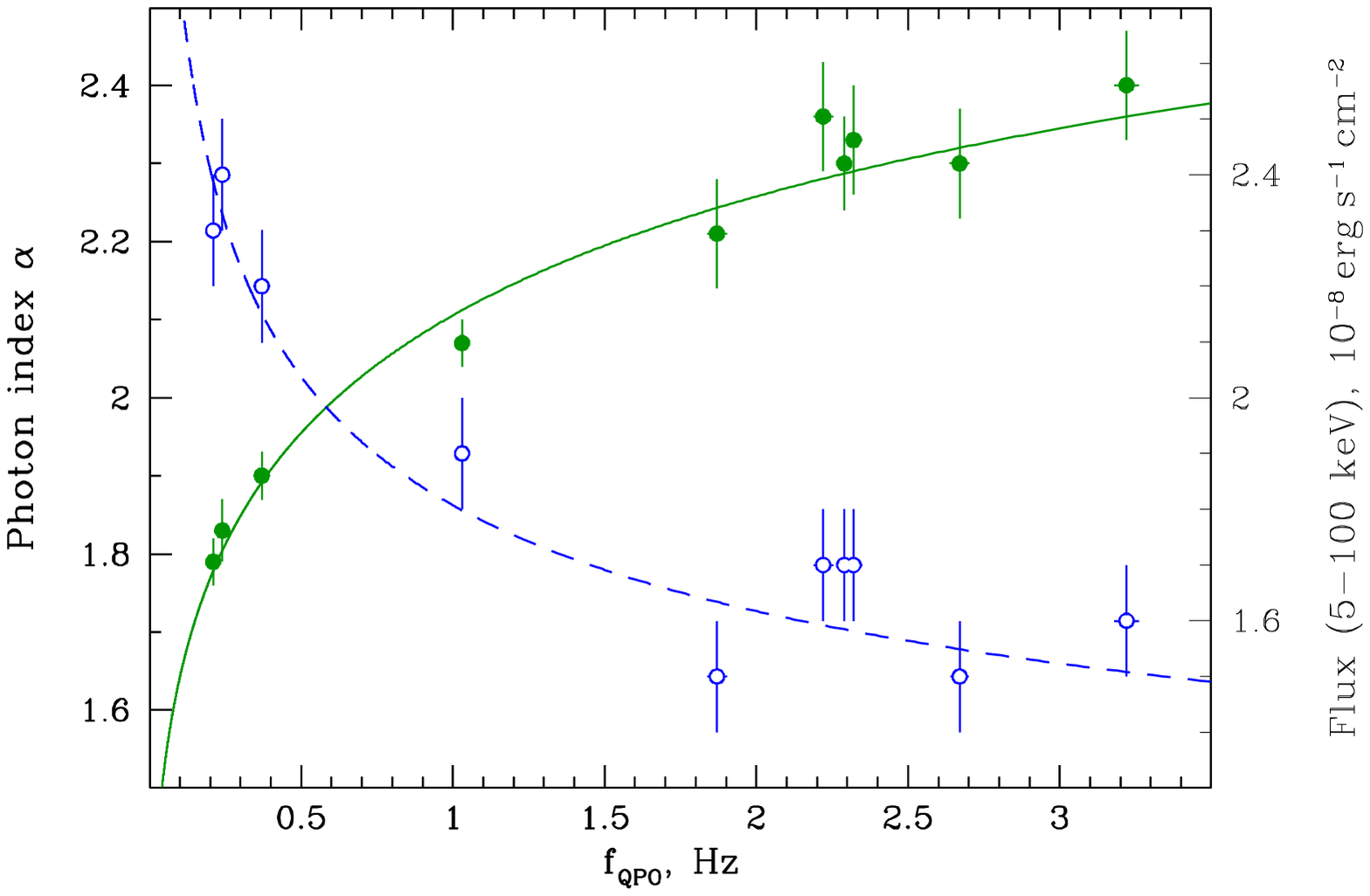}}
\caption{Photon index $\alpha$ (filled circles) and flux (open
  circles) of the broadband (5--150 keV, the MAXI and BAT/SWIFT
  data) X-ray spectrum for MAXI\,J1535-571 versus QPO frequency
  (measured by the XRT/SWIFT and IBIS/ISGRI/INTEGRAL
  telescopes). The curves indicate the best fits described in
  the Section ``Discussion''.}
\label{fig:gvsf}
\end{figure*}
%----------------------------------------------------------------------------------------------

We also used data from the IBIS/ISGRI and \mbox{JEM-X}
telescopes of the INTEGRAL observatory in searching for
QPOs. The contribution of the source to the total photon count
rate by coded-aperture telescopes is appreciably smaller than
its contribution to the count rate by focusing telescopes.
Therefore, the low-frequency noise amplitude and the overall
shape of the power spectrum for the source measured by these
telescopes differ from those measured by XRT. The energy ranges
in which the XRT and IBIS/ISGRI measurements were performed also
differ. Nevertheless, the QPO peak in the power spectra obtained
by the INTEGRAL telescopes is clearly seen, while the QPO
frequency was measured by these telescopes with a high accuracy
owing to the long exposures ($\ga15$ ks). The results of these
measurements are also included in Table\,1. Because of the
harder X-ray band, the QPO peak in these spectra is appreciably
narrower than that in the XRT/SWIFT power spectra.

As has already been mentioned, after revolution 1860 the
INTEGRAL observations were interrupted due to a solar flare, the
instruments were activated, and the background of charged
particles remained high for a long time after the resumption of
their operation. Besides, the hard X-ray flux from the source
dropped by almost a factor of 2 by this time and, therefore, the
significance of the QPO detection by the IBIS/ISGRI telescope
decreased dramatically.  However, the QPO peak remained clearly
seen in the power spectra measured by the \mbox{JEM-X} telescope
in revolutions 1862--1863 (MJD\,58009.3--58015.3), because the
soft X-ray flux from the source rose dramatically.  This allowed
the XRT/SWIFT measurements of the QPO frequency to be
supplemented at a late stage of the observing period under
discussion.
%----------------------------------------------------------------------------------------------
\begin{table*}[t]

\vspace{6mm}
\centering
{{\bf Table 2.} Parameters of the integral X-ray spectrum of MAXI\,J1535-571\a} 

\vspace{5mm}\begin{tabular}{c|c|c|c|c|c} \hline\hline
  & \multicolumn{2}{c|}{} & & \\ [-3.3mm]
Observation &Beginning &End&  QPO
&Photon
&Flux\dd\\ \cline{1-3}
                      & \multicolumn{2}{c|}{} & & \\ [-3.3mm]
SWIFT ObsID  & \multicolumn{2}{c|}{MJD--58000.00} & frequency\bb
&index, $\alpha$\cc\\
\hline
& & &&&  \\ [-3.3mm]
10264003 &4.28&4.36    &$0.42\pm 0.02 $ & $1.79\pm0.03$&$2.3\pm0.1$\\
39407002 &4.63&4.93    &$0.24\pm 0.01 $ & $1.83\pm0.04$&$2.4\pm0.1$\\
87473001 &5.28&5.45    &$0.37\pm 0.01 $ & $1.90\pm0.03$&$2.2\pm0.1$\\
30806051 &6.27&6.95    &$1.03\pm 0.01 $ & $2.07\pm0.03$&$1.9\pm0.1$\\ 
10264004 &7.27&7.29    &$1.87\pm 0.03 $ & $2.21\pm0.07$&$1.5\pm0.1$\\
10264007 &9.01&9.02    &$2.67\pm 0.03 $ & $2.30\pm0.07$&$1.5\pm0.1$\\
10264006 &10.93&10.94&$2.22\pm 0.03 $ & $2.36\pm0.07$&$1.7\pm0.1$\\
10264008 &11.39&11.40&$2.32\pm 0.03 $ & $2.33\pm0.07$&$1.7\pm0.1$\\
10264009 &12.12&12.13&$2.29\pm 0.02 $ & $2.30\pm0.06$&$1.7\pm0.1$\\
10264010 &14.18&14.19&$3.22\pm 0.04 $ & $2.40\pm0.07$&$1.6\pm0.1$\\
\hline
\multicolumn{6}{l}{}\\ [-2mm]
\multicolumn{6}{l}{\a\ The best fit to the spectrum in the
  energy range 5--150 keV from the}\\ 
\multicolumn{6}{l}{\ \ \ GSC/MAXI and BAT/SWIFT data.}\\
\multicolumn{6}{l}{\bb\ The QPO fundamental harmonic frequency
  determined from the}\\
  \multicolumn{6}{l}{\ \ \ XRT/SWIFT or IBIS/ISGRI/INTEGRAL data, Hz.}\\
\multicolumn{6}{l}{\cc\ The photon index in the energy range 5--100 keV.}\\
\multicolumn{6}{l}{\dd\ The 5--100 keV flux, $10^{-8}\ \mbox{erg~s}^{-1}$ cm$^{-2}$.}\\
\end{tabular}
\label{tab:gvsf}
\end{table*}
%----------------------------------------------------------------------------------------------

Figure~\ref{fig:qpoevol} shows how the QPO frequency changed
during the outburst. Within the first hours of observations
(MJD\,58004.3--58005.0) the QPO frequency was $\sim 0.23$ Hz,
but then it began to monotonically increase.  Between
MJD\,58009.02 and MJD\,58010.56, at the instant the break was
observed in the light curves of the X-ray nova, there was a
reverse jump by $\sim 0.7$ Hz in the QPO frequency, after which
the frequency continued to increase, but more slowly.  The
filled circles in Fig. \,\ref{fig:qpoevol} indicate the
evolution of the spectral softness $S$ in the standard X-ray
band (the ratio of the fluxes in the 2--4 and 4--10 keV
subbands). It remarkably correlates with the change in the QPO
frequency. Obviously, this parameter reflects mainly the changes
occurring in the blackbody radiation from the accretion
disk\footnote{The spectral measurements performed in this period
  by the NuSTAR observatory (Xu et al. 2017) show that the
  contribution from the photons of the blackbody component to
  the spectrum at energies $\ga4$ keV drops
  sharply. Nevertheless, the change in the QPO frequency turned
  out to correlate with the change in the 2--4 keV flux more
  poorly than with this spectral softness.}. Unfortunately, the
available data do not allow us to perform real spectral
measurements of the source's radiation in this band and to trace
the evolution of the parameters of the blackbody spectral
component (its temperature and truncation radius) in more
detail.

It is also natural to investigate the dependence of the QPO
frequency on the photon index of the hard part of the source's
radiation spectrum, in particular, to check whether a
characteristic saturation of the photon index is observed as the
QPO frequency increases (see, e.g., Sobczak et al. 2000;
Vignarca et al. 2003). For this purpose, we used the broadband
MAXI\,+\,BAT/SWIFT spectra that were fitted in the energy range
5--150 keV by an absorbed power law with an exponential cutoff
(\texttt{const*phabs*cutoffpl}).  The hydrogen absorption column
density was fixed at $N_{\rm H}=4\times10^{22}$ cm$^{-2}$; the
cross-calibration constant of the instruments in all cases
differed from unity by no more than 15\%. Unfortunately, the
short exposure of the BAT/SWIFT observations does not allow the
exponential cutoff energies to be measured accurately (the
typical error is $\simeq15$ keV for 40--60 keV) and, therefore,
$E_{\rm cut}$ was fixed at a mean value of $50$ keV.  The energy
range 2--5 keV was disregarded, because the component associated
with the blackbody disk radiation contributed noticeably to
it. The changes in photon index are presented in Table\,2 and
indicated in Fig.~\ref{fig:gvsf} by the filled (green)
circles. A gradual saturation of the photon index near
$\alpha\sim2.35$ as the QPO frequency increases above $\sim2$ Hz
seems to be actually present. The open (blue) circles in the
same figure indicate the drop in flux in the energy range 5--100
keV.
%----------------------------------------------------------------------------------------------
\begin{figure*}[th]
\includegraphics[width=\linewidth]{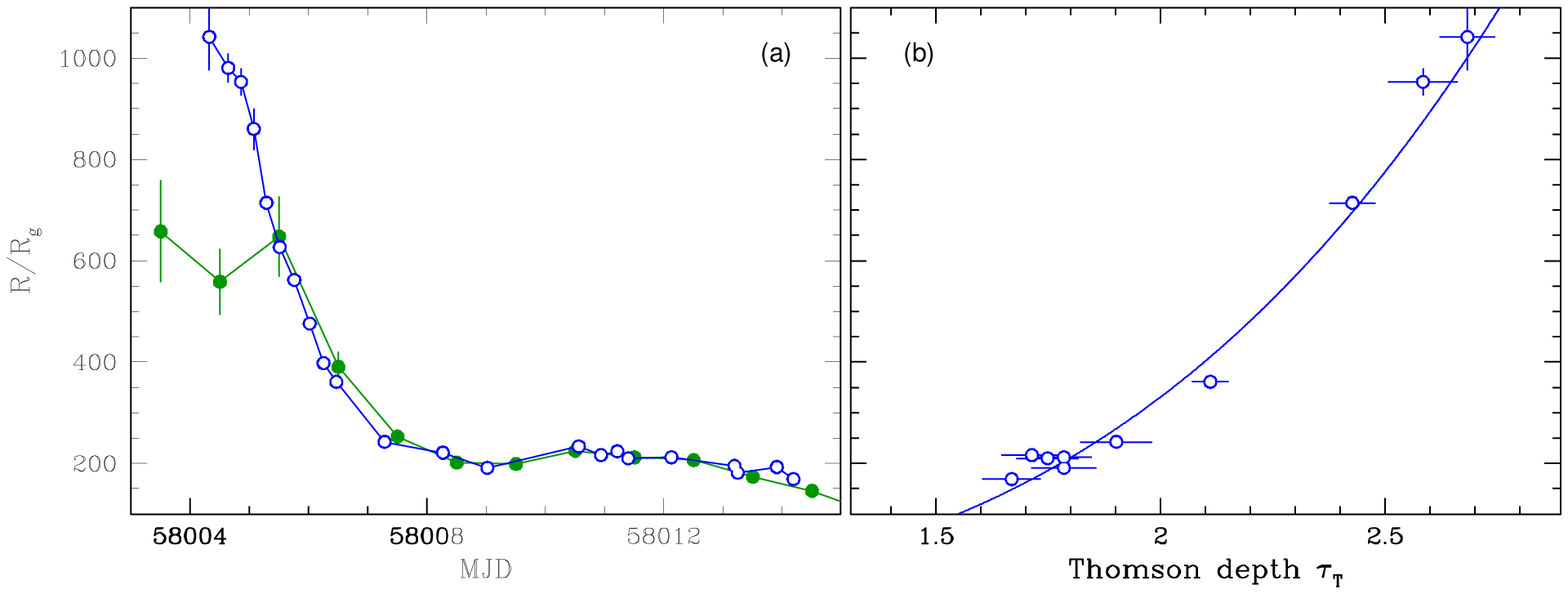}
\caption{(a) Change in the truncation radius of the blackbody
  accretion disk in the X-ray nova MAXI\,J1535-571 with time at
  the initial outburst stage. It was determined from the
  observed evolution of the LF QPO frequency in the power
  spectrum of the source (blue open circles) and the evolution
  of its X-ray spectral softness (filled green circles). \protect\\
  (b) Dependence of the Thomson depth of the central
  high-temperature accretion disk zone in the source on its
  radius (blackbody disk truncation radius). It was determined
  from the change in the photon index of the hard X-ray
  radiation.}
\label{fig:rtr}
\end{figure*}
%----------------------------------------------------------------------------------------------

%*************************************************************
\section*{DISCUSSION}
\noindent
We traced the evolution of the low-frequency QPO parameters in
the power spectrum of a new black hole candidate, the X-ray nova
MAXI\,J1535-571, during its outburst in September 2017, at the
rising phase, when it was in the hard spectral state. Using data
from three X-ray telescopes, we traced the increase in the QPO
frequency from $\sim0.2$ to $\sim3$ Hz for $\sim10$ days, until
the source's transition to the high soft state, and investigated
the correlation of the QPO frequency with the X-ray spectral
hardness.

The nature of QPOs has not yet been established.  In some models
for the origin of QPOs (see, e.g., Titarchuk et al. 2007) their
frequency is determined by the Keplerian rotation frequency of
the material in the accretion disk at its truncation radius,
i.e., the boundary of the inner high-temperature region
responsible for the observed hard X-ray radiation. Indeed, the
appearance of periodic perturbations that are subsequently
enhanced in the radiation-dominated plasma of the inner zone
might be expected in this disk region.  From the data in
Fig.\,\ref{fig:qpoevol} we can then estimate the disk truncation
radius $R_{\rm tr}=(\sqrt{2} f_{\rm QPO} R_{\rm g}/c)^{-2/3}
R_{\rm g}$ and trace its change with time. Here, $R_{\rm
  g}=2GM/c^2$ is the gravitational radius of a black
hole. The result of such a calculation by assuming the black
hole mass to be $M=10\ M_{\odot}$ is indicated in
Fig.\,\ref{fig:rtr} (left panel) by the open (blue) circles. The
behavior of the truncation radius is characterized by a rapid
decrease from $\sim10^3\ R_{\rm g}$ to $\sim200\ R_{\rm g}$
within the first three days after MJD\,58005, a succeeding
five-day period of relative stability, and a very slow further
decrease to $\sim160\ R_{\rm g}$ in the last two days.

The behavior of the truncation radius can be estimated
independently from the law of change in spectral softness $S$
presented in Fig.\,\ref{fig:qpoevol} by taking into account the
surprising correlation of this parameter with the QPO frequency
and by assuming its proportionality to the surface temperature
of the blackbody disk near its inner edge. According to Shakura
and Sunyaev (1973), the dependence of this temperature on
$R_{\rm tr}$ is given by the expression $\sigma T_{\rm
  tr}^4=(3/8\pi)\ GM\dot{M}/R_{\rm tr}^3$, where $\dot{M}$ is
the accretion rate and $\sigma$ is the Boltzmann constant; thus,
$R_{\rm tr}\sim T_{\rm tr}^{-4/3}$.  Accordingly, we attempted
to fit the truncation radius by the formula $R_{\rm
  tr}=A\,(S-S_0)^{-4/3}\ R_{\rm g}$. The best agreement was
achieved at $A=108\pm2$ and $S_0=0.60\pm0.01$. The best fit is
indicated in Fig.\,\ref{fig:rtr} (left panel) by the filled
(green) circles. The two estimates agree well between themselves
at $R_{\rm tr}\la 600\ R_{\rm g}$. At larger truncation radii
the disk temperature is apparently too low to be adequately
described by the parameter $S$ determined at $h\nu>2$ keV.

On the other hand, interesting estimates of the parameters of
the inner hot disk zone can be obtained from the spectral
variability of the hard X-ray radiation. For example, by
considering the inner hot disk zone as a quasi-spherical one and
by assuming that its electron temperature $kT_{\rm e}\simeq50$
keV does not depend on the radius, we can use the well-known
formula of the Comptonization theory (Sunyaev and Titarchuk
1980) for the photon index of the emission forming in the cloud:
$$\alpha=-\frac{1}{2}+\sqrt{\frac{9}{4}+\gamma},$$ where
$\gamma=(\pi^2/3)(\tau_{\rm T}+2/3)^{-2}\ (m_{\rm e}c^2/kT_{\rm
  e}),$ and estimate the Thomson optical depth of the cloud (see
Fig.\,\ref{fig:rtr}, right panel). In this figure the optical
depth $\tau_{\rm T}$ is presented as a function of the radius of
the inner zone $R_{\rm tr}$, which we found by again assuming
that the QPOs forming here have a Keplerian frequency.  The
optical depth $\tau_{\rm T}$ changes from $\sim2.7$ to
$\sim1.5$, in agreement with our assumptions. We see that
$\tau_{\rm T}$ and $R_{\rm tr}$ correlate well (unambiguously)
between themselves.  We fitted this dependence by the formula
$\tau_{\rm T}=(R_{\rm tr}/R_*+B)^\beta$. The best agreement was
achieved at $B=1.8\pm0.1$, $R_*=(40\pm2)\,R_{\rm g}$, and
$\beta=0.3\pm0.01$; the best fit is indicated in
Fig.\,\ref{fig:rtr} (right panel) by the solid line. The solid
curve in Fig.\,\ref{fig:gvsf} indicates the result of using this
formula to estimate the photon index $\alpha$ within the
Comptonization theory. We see good agreement of this curve with
the results of direct $\alpha$ measurements.

This formula describes the observed drop in the hard
(Comptonized) X-ray flux from the source in this period equally
well. If the radiation originates in the quasi-spherical cloud
of an optically semitransparent plasma, then its flux at other
constant cloud parameters must be directly proportional to its
Thomson optical depth (Sunyaev and Titarchuk 1980). Indeed, our
attempt to fit the drop in flux indicated in
Fig.\,\ref{fig:gvsf} by the open (blue) circles by the formula
$F=D\,\tau_{\rm T}$ turned out to be successful --- the result
for the best value of the parameter $D=0.88\pm0.01$ is indicated
in this figure by the dashed (blue) line.
%----------------------------------------------------------------------------------------------
\begin{figure}[ht]
\centerline{\includegraphics[width=1.1\linewidth]{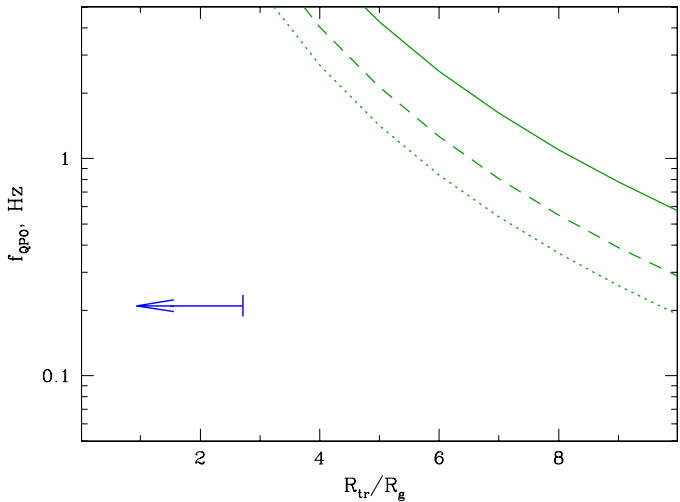}}
\caption{LF QPO frequency versus truncation radius according to
  the Lense-Thirring model (Ingram et al. 2014) for different
  black hole masses $10$, $20$, and $30\ M_{\odot}$ (the solid,
  dashed, and dotted lines, respectively). The black hole spin
  is taken to be 0.84 (the lower limit obtained by Xu et
  al. (2017) from the NuSTAR satellite data near
  MJD\,58004.3). The arrow indicates the maximum truncation
  radius from the NuSTAR data, the QPO frequency was measured
  based on the simultaneous XRT/SWIFT observations.}
\label{fig:qpocons}
\end{figure}
%----------------------------------------------------------------------------------------------

On the whole, it can be said that the assumption that the QPO
frequency is equal to the Keplerian frequency at some radius
comparable to the truncation radius of the blackbody accretion
disk in MAXI\,J1535-571 leads to self-consistent and reasonable
results. However, this contradicts the results from Xu et
al. (2017) of a spectral analysis of the NuSTAR data obtained
during the source's observation almost simultaneously with our
first detection of QPOs (at a frequency of $\sim0.21$ Hz in the
interval MJD\,58004.28--58004.36). The authors obtained
stringent constraints on the black hole spin ($a>0.84$) and the
disk truncation radius ($R_{\rm tr}<2.01\,R_{\rm ISCO}\simeq
2.7\ R_{\rm g}$). At such a truncation radius the Keplerian
frequency is much higher than the measured QPO frequency. The
only known QPO model that could give a reasonable QPO frequency
in this case is the model by Ingram et al. (2009, 2014) for the
formation of QPOs through the Lense-Thirring precession of the
inner hot disk region.  According to Ingram et al. (2014),
depending on the black hole mass $M$ and spin $a$, the QPO frequency
is defined by the formula
$$
\frac{f_{\rm QPO}}{f^*_{\rm K}}=
\left(1-\sqrt{1-\sqrt{2}a\,r^{3/2}+0.75 a^2 r^{2}}\ \right),
$$ where $r=R_{\rm g}/R_{\rm tr}$ and the relativistic Keplerian
frequency
$$
f^*_{\rm K}=\left(\frac{c}{\pi R_{\rm g}}\right) \left[\left( \frac{2}{r}\right)^{3/2}+a\right]^{-1}.
$$
The predictions of the Lense-Thirring model calculated
with this formula for a black hole with a mass of
$10,\ 20,$ and even $30\ M_{\odot}$, which are presented in Fig.~\ref{fig:qpocons},
differ greatly from the observations. This contradiction
can arise both from the incorrect interpretation
of the spectra and from the inconsistency of the QPO
generation mechanism with the proposed one.

%*************************************************************
\section*{ACKNOWLEDGMENTS}
\noindent
This work is based on the data from the INTEGRAL observatory
retrieved via its Russian and European Science Data Centers, the
SWIFT observatory retrieved via NASA/HEASARC, and the MAXI
experiment onboard the International Space Station (ISS)
retrieved via the site \texttt{maxi.riken.jp/top}. We are
grateful to the Russian Foundation for Basic Research (project
no. \mbox{17-02-01079-a}), the program of the President of the Russian
Federation for support of leading scientific schools (project
no. \mbox{NSh-10222.2016.2}), and the ``Transitional and Explosive
Processes in Astrophysics'' subprogram of program P-7 of the
Presidium of the Russian Academy of Sciences for the financial
support.
%*************************************************************
\begin{appendix}

  \section*{EXTRACTION OF THE BAT/SWIFT SPECTRA}	

\noindent
The BAT/SWIFT telescope can operate in several different modes;
some modes can be used simultaneously. The survey mode is used
in most observations: in this mode the telescope is pointed at
the chosen object, while the data acquisition system writes the
distribution of recorded photons in energy for each detector
pixel over the entire exposure time (usually from 500 to
1500~s). Since BAT/SWIFT at each instant sees approximately one
sixth of the sky, using the survey data allows the spectra of
many bright ($\ga100$ mCrab in the 15--150 keV band) sources to be
regularly obtained. We used the following data processing
procedure:
%\begin{enumerate}

%\item
(1) For each such histogram we selected the working and
  healthy detector pixels (\texttt{batdetmask}, here and below,
  the commands from the \texttt{Heasoft} package are given in
  parentheses). Then, we constructed an image of the detector
  plane (\texttt{batbinevt}) and searched for ``hot'' pixels
  (\texttt{bathotpix}).

%\item
(2) We constructed an image of the detector plane
  (\texttt{batbinevt}) once again, eliminating the previously
  found nonworking or hot pixels. A sky image and a partial
  coding map were constructed (\texttt{batfftimage}). Sources
  were recorded in the sky image (\texttt{batcelldetect}).

%\item
(3) If a source was recorded at a statistically significant
  level, then the input data were subjected to a more accurate
  recalibration (\texttt{baterebin}). We constructed a model
  shadow image of the source in the detector plane
  (\texttt{batmaskwtimg}) and extracted its spectrum
  (\texttt{batbinevt}).

  %\item
(4) The spectrum was corrected by taking into account the
  ray-tracing results (\texttt{batupdatephakw}). The errors of
  the fluxes in the energy channels were increased by taking
  into account the experimentally determined systematic errors
  (\texttt{batphasyserr}); the telescope response matrix was
  calculated (\texttt{batdrmgen}).
%\end{enumerate}

The spectra obtained in this way were used to construct the
source's broadband spectra.
\end{appendix}

%****************************************************************
\label{lastpage}

\bibliographystyle{pazh} \bibliography{reflist_rus}

\vspace{5mm}

\begin{flushright}
  {\sl Translated by V. Astakhov}
\end{flushright}
\end{document}